\documentclass[runningheads]{llncs}

\usepackage{graphicx}
\usepackage{CJK}

\begin{document}

\title{How People Perceive The Dynamic Zero-COVID Policy: A Retrospective Analysis From The Perspective of Appraisal Theory}
\titlerunning{How People Perceive The Dynamic Zero-COVID Policy}

\author{Na Yang\inst{1} \and
Kyrie Zhixuan Zhou\inst{2} \and 
Yunzhe Li\inst{3}}
\authorrunning{Yang et al.}
\institute{
Wuhan University, Wuhan, China \\
\email{2022201020044@whu.edu.cn}\\
\and
University of Illinois at Urbana-Champaign, USA\\
\email{zz78@illinois.edu}
\and
Shanghai Jiao Tong University, China\\
\email{yunzhe.li@sjtu.edu.cn}
}

\maketitle              

\begin{abstract}
The Dynamic Zero-COVID Policy in China spanned three years and diverse emotional responses have been observed at different times. In this paper, we retrospectively analyzed public sentiments and perceptions of the policy, especially regarding how they evolved over time, and how they related to people's lived experiences. Through sentiment analysis of 2,358 collected Weibo posts, we identified four representative points, i.e., policy initialization, sharp sentiment change, lowest sentiment score, and policy termination, for an in-depth discourse analysis through the lens of appraisal theory. In the end, we reflected on the evolving public sentiments toward the Dynamic Zero-COVID Policy and proposed implications for effective epidemic prevention and control measures for future crises.

\end{abstract}

\section{Introduction}

Since the outbreak of COVID-19, stringent policies and measures have been implemented to contain the pandemic in many countries, such as stay-at-home policy, lockdowns, school and workplace closures, etc. \cite{allen2022covid}. The Chinese government has adhered to the Dynamic Zero-COVID Policy, to keep COVID cases as close to zero as possible \cite{bai2022optimizing}. To achieve this, it has implemented mass testing, quarantined infected people in government facilities, and imposed citywide lockdowns \cite{tan2023dynamic}. Understanding how such anti-epidemic policies and measures worked and how they affected people's lives has drawn much attention from the research community \cite{liu2022dynamic,shi2023ethical,zhai2023lockdown,zhan2023zero}.

While people's sentiments and perceptions of public health emergencies as well as corresponding administrative policies are of vital importance to the improvement of emergency responses \cite{zhang2020effective}, little research has been devoted to providing a comprehensive picture of public opinions for the Dynamic Zero-COVID Policy in China, especially how they changed through the course of the pandemic. Existing research mostly focused on public opinions during the early stage of the COVID outbreak \cite{shi2022online}. We bridge this research gap with the current study.

To provide a comprehensive understanding of public opinions toward the Dynamic Zero-COVID Policy and how they evolve, we constructed a dataset of Weibo\footnote{Weibo, a.k.a., Sina Weibo, is a Chinese microblogging website.} discussions spanning three years. Notably, social media platforms have become the main way of communication during COVID in China \cite{liao2020public}. People relied heavily on Weibo to express their opinions and share information \cite{zhao2020chinese}. Thus, our analysis drew on people's online discussions on Weibo to understand public opinions and sentiments. 

We first conducted a sentiment analysis to identify the trend of sentiment evolvement and outstanding points (i.e., outliers). We then identified several representative points for an in-depth discourse analysis with appraisal theory \cite{Martin2005TheLO}, including (1) policy initiation where a high peak of sentiment was seen, (2) policy termination with a positive sentiment, (3) a sudden sentiment change, and (4) Shanghai lockdown which led to a low sentiment peak. 

The appraisal framework allows both the manifestation of emotionality across discourse and the in-depth analysis of the linguistic resource that is used to construe the value position. Our discourse analysis revealed that negative words reflecting people’s fear and insecurity were commonly used at the onset of the pandemic, while feelings of relief and expectation for a normal life were expressed after the Dynamic Zero-COVID Policy was lifted. An emotional shift was noticed in March 2021 when many posters repeatedly used negative judgment of the propriety of certain policies. We noticed that during the Shanghai lockdown negative adjectives along with upscaled intensification were applied to convey the depression and anxiety experienced by the public.

Our contribution is thus three-fold. First, we constructed a dataset of online discussions around COVID policy, which could spur more research in understanding public opinions and people's lived experiences during public health crises. Second, we used a mixed-methods analysis to unveil the dynamic attitudinal and perceptual shifts and the underlying reasons during the three years of COVID-19 prevention and control. Third, our results and proposed implications can provide references for future policymaking when another public health emergency happens. 
\section{Related Work: Public Opinions Toward COVID Policies}

Since the outbreak of the COVID-19 pandemic, governments around the world have put forward various policies to curb the spread of the epidemic, such as social distancing, lockdown, quarantine, etc. \cite{bai2022optimizing}. 

Public opinions and attitudes toward epidemic prevention and control measures are frequently studied. 
Gadarian et al. investigated the U.S. public's health behaviors, attitudes, and policy opinions about COVID-19 in the earliest weeks of the national health crisis, revealing the central role of partisanship in shaping individual responses \cite{gadarian2021partisanship}.
Hossain et al. assessed public university students' COVID-19 vaccine-related knowledge, perceptions, attitudes, and acceptability of a COVID-19 vaccine as well as the factors affecting their perceptions using a cross-sectional e-survey. Positive attitudes toward the COVID-19 vaccine were found, and 78\% of students had adequate COVID-19 vaccine-related knowledge \cite{hossain2021covid}. Purnama and Susanna et al. conducted a cross-sectional study to investigate the variables that influenced attitudes toward large-scale social restriction policies in Indonesia \cite{purnama2020attitude}. The results showed that people supporting this policy had a positive influence on the successful implementation of this program and public attitudes were affected by the perceived benefits, negative and positive perceptions, and the threat associated with COVID-19.

Besides surveys, social media analysis \cite{rim2020polarized,baum2019media} has also been adopted to understand public opinions towards COVID policies. For example, Tsai and Wang conducted a sentiment analysis of Twitter data and showed a strong correlation between sentiment changes in COVID-19-related Twitter data and public health policies and events \cite{tsai2021analyzing}. By taking a cultural sensitivity perspective and collecting data on both Twitter and Weibo, Luo et al. conducted a public perception comparison around the COVID-19 vaccine using semantic network analysis \cite{luo2021exploring}. The findings suggested that these two groups of social media users shared overlapping themes in their discussion, including domestic vaccination policies, priority groups, challenges from COVID-19 variants, and the global pandemic situation.

Public attitudes and opinions toward epidemic prevention and control measures can vary based on people's demographic background as well as political and cultural situations in a certain country/region. 
Glanz et al. found that there were significant regional and political differences in people's reactions to US federal government policies \cite{glanz2022correlates}.
Sabat et al. surveyed people in seven European countries to address people's support for containment policies, worries about COVID-19 consequences, and trust in sources of information and found a north-south divide in public opinions \cite{sabat2020united}. 
In addition, there are significant differences between males and females in their self-risk perception \cite{stephen2023gender}.

People from China showed a high level of support for COVID policies in prior research. Song et al. conducted an online survey to examine the attitudes of people from Hubei province, who experienced the pandemic at the earliest stage, towards quarantine at the height of the COVID-19 outbreak \cite{song2020public}. The results indicated that the public strongly supported quarantine as a necessary measure to control the pandemic. Xu et al. used an online questionnaire to explore residents' knowledge, attitudes, and practice behaviors during the outbreak of COVID-19 and found that 85.5\% responded positively toward the mandatory public health interventions implemented nationwide by the Chinese authorities \cite{xu2021public}. 

\textbf{Research gaps.}
Existing research concerning public opinions towards epidemic prevention and control measures are mainly cross-sectional studies that observe and analyze data at a single point in time. Nevertheless, public opinions are never static but evolve with constant changes in situation and policy. China maintained and adjusted the Dynamic Zero-COVID Policy for three years, providing an ideal site for investigating how public opinions evolved. Based on this consideration, we aspire to understand the evolving public opinions over a longer period, i.e., from policy initiation to policy termination, with the current study.
\section{Methodology}
By combining an overall sentiment analysis and an in-depth discourse analysis, the present study explored the micro characteristics and patterns of discourse under the premise of macro trends. The overall sentiment trend revealed the dynamic process of attitudinal shifts over a period of time, based on which a more specific discourse analysis was conducted on key points selected by researchers. 

\subsection{Data Collection}
To portray a full picture of people's attitudinal shifts in three years of epidemic prevention and control measures in China, the data of this study were collected from December 31, 2019, to December 28, 2022, using the keywords “Dynamic Zero-COVID Policy, epidemic prevention, pneumonia, COVID, and pandemic," in Chinese, on Weibo. A constructed week sampling was adopted for its accuracy of reflecting the periodic changes on a weekly basis \cite{luke2011much}. Altogether 79 days were selected with 11 complete constructed weeks. Thirty posts were randomly selected per day, resulting in a dataset with 2,358 posts. Notably, only 18 posts were found on December 31, 2019, possibly due to the censorship of COVID and politics-related discussions on Weibo \cite{chen2023we}. 
Three researchers independently collected a part of the dataset and reviewed the whole dataset to ensure the relevance of the posts.

\subsection{Data Analysis}
\subsubsection{Sentiment Analysis}
Sentiment analysis is used to classify people's attitudes towards a certain topic into various categories, such as positive, negative, and neutral \cite{review_sentiment}. We calculated sentiment scores for Weibo posts using Cemotion \cite{cem}, a BERT-based \cite{bert} Python library tailored for sentiment analysis in Chinese. Essentially, Cemotion functions by analyzing and interpreting the sentiment intrinsic of written materials. It quantitatively evaluates sentiment in text and assigns a sentiment score within the interval [0, 1]. A sentiment score close to 0 is indicative of a negative sentiment, whereas a score approaching 1 signifies a positive sentiment.

\subsubsection{Discourse Analysis}
While sentiment analysis is not able to interpret details of text and interpersonal meanings behind the discourse, we further apply a discourse analysis \cite{johnstone2017discourse} based on Appraisal Theory \cite{Martin2005TheLO} to delve into discussions at key points. Appraisal has to do with the ways that writers or speakers (posters in our case) express approval or disapproval for things, people, behavior, or ideas \cite{oteiza2017appraisal}. The appraisal framework can help achieve a richer understanding of the patterns of the interpersonal meanings beyond the manifestation of emotionality across discourse \cite{oteiza2017appraisal}.
\section{Results}
By generating the overall sentiment trend based on the collected Weibo posts, we intended to exhibit a clear picture of the sentiment change during the three years of epidemic prevention and control. Utilizing the outcomes of the sentiment analysis, a comprehensive discourse analysis was carried out on the key points, helping to understand how the public’s perceptions and opinions evolved in response to the changing circumstances.

\subsection{Sentiment Analysis}

\begin{figure}
\centering
\includegraphics[width=\textwidth]{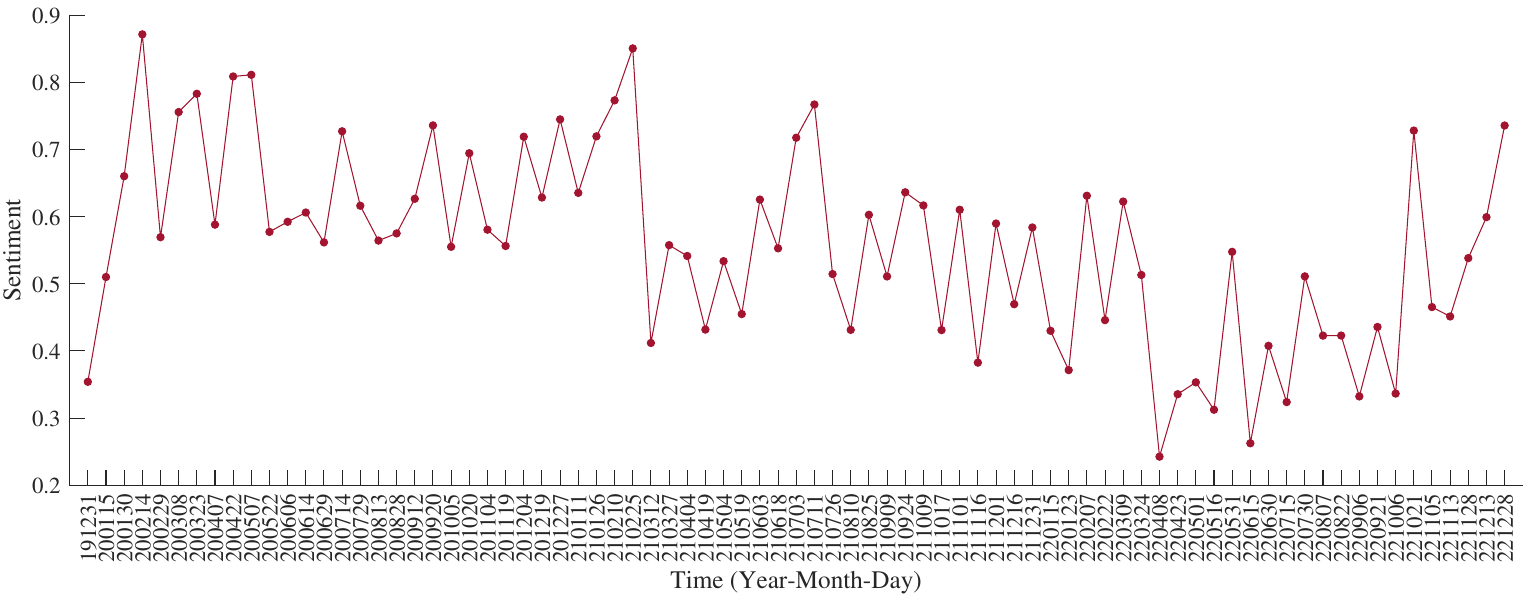}
\caption{Sentiment analysis of Weibo posts discussing the Dynamic Zero-COVID Policy. A score approaching 1 signifies a positive sentiment.} \label{fig:sentiment}
\end{figure}

The results of the sentiment analysis of COVID-related posts across three years are illustrated in Figure \ref{fig:sentiment}. When the COVID-19 pandemic started, the sentiment in Weibo posts was largely negative. Shortly after, people's sentiments became more positive, which might be related to the effectiveness of the government's strict epidemic prevention and control measures. However, over time, people's sentiments gradually declined. April 2022 saw the lowest sentiment score since the start of the pandemic when a large-scale lockdown was implemented in Shanghai. After this point, people's sentiments started to rise till December 2022, by which time the Dynamic Zero-COVID Policy was terminated. 

\subsection{Discourse Analysis}
We selected four key turning points or outliers in terms of public sentiments, i.e., policy initialization, significant sentiment change, lowest sentiment score, and policy termination, for a more in-depth discourse analysis. Note that we looked at posts in neighboring days around the selected points. 

\subsubsection{Policy Initialization.}
COVID-related discussion on Weibo started on December 31, 2019, when the pandemic broke out in Wuhan, China. People's sentiment towards COVID-19 was generally negative. The following are extracts selected from Weibo posts on December 31, 2019, and January 1, 2020. 

\textbf{Extract 1}
Many experts from Wuhan medical institutions say [Engagement: Heterogloss-dialogic expansion-attribute] the origin of the virus is unclear [-Attitude: Appreciation-composition-complexity] and it cannot be certain [Engagement: Heterogloss-dialogic expansion-entertain] that the virus is SARS. It is more [Graduation: Force-intensification] likely [Engagement: Heterogloss-dialogic expansion-entertain] to be another severe [-Attitude: Appreciation-react-ion-quality] pneumonia. Even if it is SARS, there is already a mature [+Attitude: judgment-social esteem-capacity] prevention and treatment system, so the public does not need to panic [+Attitude: Affect-security].

In Extract 1, through direct quotation, the evaluative proposition of the origin of the virus was attributed to medical experts. Although the direct quotation of reported speech could make the utterance more faithful and convincing \cite{Bell1991TheLO}, the use of dialogic expansion made room for dialogically alternative positions \cite{Martin2005TheLO}. Sentence patterns like ``it cannot be certain'' and ``it is more likely'' were employed with the effect of ``entertaining'' dialogic alternatives and reducing the interpersonal cost of doing so. The use of ``more'' indicated the possibility of other origins of the virus. The uncertainty of the virus's origin gave rise to the negative appreciation of people.
The SARS virus had a great impact on people’s lives in the past \cite{contini2020novel}, therefore people were afraid that the current virus would strongly affect people's lives again. However, people were encouraged to face the unknown with an optimistic attitude given the trustworthy prevention and treatment system established by the government.

\textbf{Extract 2} On December 31, the civil servants of sub-district office said [Engagement: Heterogloss-dialogic expansion-attribute] that the pneumonia was non-infectious [-Attitude: Appreciation-composition-complexity] and the Wuhan Huanan Seafood Wholesale Market would open as usual. But [Engagement: Heterogloss-dialogic contraction-disclaim] the market was closed the next day. How can we possibly [Graduation: Force-intensification] calm down [-Attitude: Affect-security]?

As in Extract 1, the direct quotation of the civil servants of the sub-district office in Extract 2 demonstrated the government’s evaluation of the complexity of the virus --- they thought the virus was not infectious and would not spread. In addition, they also claimed that the Huanan Seafood Wholesale Market would not be influenced by the outbreak of the virus. Although the quotation of the government officials enhanced the credibility of the utterance, the use of dialogic expansion left space for other opinions. That the situation the next day was contrary to the expectation deepened the insecurity of the public, which was conveyed by using the connective ``but'' and the adverb ``possibly.''

\textbf{Extract 3} The public is still [Engagement: Heterogloss-dialogic contraction-disclaim] fearful [-Attitude: Affect-security] of the SARS virus and the slaughter of live poultry is strictly [Graduation: Force-intensification] prohibited in the market. When did the dying embers flare up? Pandora’s Box [-Attitude: Affect-security] has been reopened.

The use of ``still'' in Extract 3 construed the emergence of the unknown virus as counter-expectational \cite{Martin2005TheLO} since people were still in the shadow of the SARS virus. The negative affect towards SARS (``fearful'') reflected people's emotional insecurity about the unknown virus. An intensification of the prohibition of the slaughter of live poultry in the market showed that people were cautious about preventing the spread of the SARS-like virus, indirectly indicating the people's negative affect of the unknown virus. ``Pandora's Box'' was a lexical metaphor that revealed people's panic about the unknown virus.

\textbf{Extract 4} I have resisted [-Attitude: Affect-desire] going out because of the pneumonia in Wuhan. Besides, the students in school all [Graduation: Force-quantification] wear masks now. 

\textbf{Extract 5} I went out today and saw many [Graduation: Force-quantification] people wearing masks. I also need to buy more [Graduation: Force-quantification] masks.

Extracts 4 and 5 showed the impact of the virus on people's lives. As Extract 4 indicated, the unknown pneumonia affected people's desire to go outside (``resist'') out of the concern of being infected with the virus. To protect themselves from the virus, many people chose to wear masks when going out. The words ``all'' in Extract 4 and ``many'' in Extract 5 added weight to the quantification of the number of people who wore masks. The use of ``more'' in Extract 5 represented the increasing demand for masks, which indirectly reflected the anxiety and insecurity of the public.

\textbf{Summary} At the time of the COVID outbreak, though experts assured people that the virus was controllable and the epidemic prevention system was mature, people were still fearful of the SARS-like virus and chose to not go out and store masks at home.

\subsubsection{Significant Emotion Change.}

A striking sentiment shift was observed on March 12, 2021. The public sentiment before that was mostly positive while negative appreciation played a major role in setting up the attitudinal stance afterwards. Below, we present extracts before and after the sudden sentiment shift. 

\textbf{Extract 1} Hungary starts using COVID-19 vaccines produced by the Chinese laboratory. I feel proud [+Attitude: Affect-happy] of my country. Many thanks to [+Attitude: Affect-satisfaction] medical workers! 

\textbf{Extract 2} A recombinant vaccine against COVID-19 developed by Wei Chen's research group is approved for clinical trials. Hopefully [+Attitude: Affect-desire] this vaccine will be available soon. I'm praying [+Attitude: Affect-desire] for an early end to the epidemic and wishing [+Attitude: Affect-desire] my country, my family, and friends all the best.

Extracts 1 and 2 were posted on February 25, two weeks before March 12. The attitudes conveyed in these extracts were positive affect towards epidemic prevention and control measures as well as vaccine development efforts. In Extract 1, the poster expressed their pride in vaccines developed by Chinese researchers and gratitude for medical workers. ``Pray'' and ``wishing'' reflected people's hope and expectation for the termination of the epidemic.

\textbf{Extract 3} 
There were no [Engagement: Heterogloss-dialogic contraction-disclaim] announcements when I bought the high-speed railway tickets, but I wasn't allowed to take the train on the day of departure [-Attitude: Affect-satisfaction].
I'm supportive [+Attitude: judgment-social esteem-tenacity] of the Dynamic Zero-COVID Policy, but the government should inform us of the details transparently [-Attitude: Affect-satisfaction]. The lack of coordination [-Attitude: Appreciation-composition-balance] among different departments has also brought inconvenience [-Attitude: Affect-satisfaction] to the public.

\textbf{Extract 4} It's unreasonable [-Attitude: judgment-social sanction-propriety] to shut down buses when there are people infected with COVID-19. The government shouldn't let the public bear the responsibility. It makes us disappointed [-Attitude: Affect-satisfaction]. 

People have long been restricted in all aspects of life because of epidemic prevention and control measures \cite{lau2022measurement} and negative emotions broke out around the sudden shift of sentiment. Inconvenience and obstacles arose in the process of implementing the strict measures. Extracts 3 and 4 reflected people's complaints about the inconvenience of traveling caused by unreasonable policies, which completely stopped public transportation. The use of ``no'' meant that some dialogic alternative was represented as not applying \cite{Martin2005TheLO}. As Extract 3 exemplifies, there were no announcements concerning the epidemic prevention and control measures when people bought railway tickets. The emotion of unhappiness and dissatisfaction was displayed through negative appreciation of the lack of coordination between different government apartments (``the lack of coordination''). 
Public transportation like buses are confined spaces, which were deemed easier for the virus to transmit. Therefore, some local governments would take temporary traffic control measures to suspend the buses. 
In Extract 4, the author used ``unreasonable'' to demonstrate their negative judgment of the propriety of such policy. Dissatisfaction was exhibited through the use of the adjective ``disappointed.''

\textbf{Extract 5} Negative COVID-19 results can be used for seven days. Why do the accompanying family members of the patients have to take a new COVID test when they leave the hospital, but [Engagement: Heterogloss-dialogic contraction-disclaim] the doctors don't [-Attitude: judgment-social sanction-propriety]? I wonder if [Engagement: Heterogloss-dialogic expansion-entertain] it is an effective anti-epidemic measure if it only restricts the accompanying family members of the patients [-Attitude: judgment-social sanction-propriety]. 

\textbf{Extract 6} I skimmed through the school notices and found that most of the regulations are reasonable [+Attitude: judgment-social sanction-propriety]. But [Engagement: Heterogloss-dialogic contraction-disclaim] why can't students leave campus while teachers can enter and leave schools freely [-Attitude: judg-ment-social sanction-propriety]?

Extracts 5 and 6 concerned public attitudes towards the varying treatment for different groups by the anti-epidemic measures. Stricter rules were applied to certain groups of people. Extract 5 revealed the poster's doubt and negative attitude towards the unreasonable regulation posed by the hospital -- the patients' family members were required to take a new COVID test if they left the hospital, which did not apply to medical workers. Since appraisal involves the negotiation of solidarity \cite{Martin2005TheLO}, the poster used the sentence pattern ``I wonder if'' to invite empathy from the audience and construe a heteroglossic backdrop for the text so as to suggest that the particular value position was actually in tension with dialogistic alternatives, implying the negative judgment of the propriety of the policy. In Extract 6, the poster first showed their positive judgment of the propriety of most school policies and expressed their doubt of one policy which only restricted the autonomy of the students and did not apply to teachers. Such double standards triggered dissatisfaction.

\textbf{Summary} Long-lasting restrictions on people's traveling and life eventually caused a sharp attitudinal change. People were dissatisfied with the shutdown of public transportation, which was regarded by the government as an effective anti-epidemic measure but perceived by the public as unreasonable. The policies that demonstrated differential treatment for different social groups were criticized by people, which they thought were unreasonable and did not make sense scientifically.

\subsubsection{Lowest Sentiment Score --- Shanghai Lockdown}
\label{shanghai}

% \rev{third person}

Since the lockdown of Shanghai on March 28, 2022, the lowest sentiments were seen on April 8, 2022. Negative attitudes toward the strict epidemic prevention and control measures were expressed by people both in and outside Shanghai. Positive attitudes were found in people outside Shanghai in our sampled data during the lockdown. 

\textbf{Extract 1} For those who are not [Engagement: Heterogloss-dialogic contrac-tion-disclaim] in good health conditions and do not [Engagement: Heterogloss-dialogic contraction-disclaim] receive any help, they'll just remain hungry. It's ridiculous [-Attitude: judgment-social esteem-normality] that people are starving in a peaceful [-Attitude: Appreciation-reaction-quality] age.

\textbf{Extract 2} Living in Shanghai is really [Graduation: Force-intensification] scary [-Attitude: Affect-security]. Severely [Graduation: Force-intensification] ill older adults cannot [Engagement: Heterogloss-dialogic contraction-disclaim] be treated in time. Young people live without food for three days. There are no [Engagement: Heterogloss-dialogic contraction-disclaim] supplies at home, and government supplies cannot [Engagement: Heterogloss-dialogic contraction-disclaim] be delivered to home. I'm so [Graduation: Force-intensification] lost [-Attitude: Affect-security]. 

Extracts 1 and 2 demonstrated a severe lack of supplies in Shanghai. The poster for Extract 1 used ``ridiculous'' to describe the abnormality of the phenomenon that people were starving. Such abnormality was strengthened by the positive appreciation of the current ``peaceful age.'' The adjective ``scary'' in Extract 2 reflected the insecurity of the public, and it was enhanced by upscaled intensification (``really''). The repeated usage of disclaim highlighted the severe situation in Shanghai. The poster stressed their confusion and depression by using ``so lost.''

\textbf{Extract 3} Everyone is vaccinated. There's not [Engagement: Heterogloss-dialogic contraction-disclaim] a single [Graduation: Force-quantification] Omic-ron-related death. But [Engagement: Heterogloss-dialogic contraction-disclaim] what about the other problems? People don't [Engagement: Heterogloss-dialogic contraction-disclaim] have food. Newborns don't [Engagement: Heterogloss-dialo-gic contraction-disclaim] have milk powder. The elderly don't [Engagement: Heterogloss-dialogic contraction-disclaim] have medicines. Is it really [Graduation: Force-intensification] necessary to let people suffer [-Attitude: Affect-satisfaction], not because of the epidemic, but because of the anti-epidemic measures? People in Shanghai are also scolded [-Attitude: Affect-happy] by people from other provinces.

The repeated employment of disclaim in Extract 3 portrayed a picture of people's hard lives during the lockdown. The adverb of degree ``really'' was applied to intensify the people's rational criticism of the COVID policy. They thought people were suffering because of the policy instead of the epidemic itself. The words ``suffer'' and ``scold'' revealed the negative affect of the public. 

\textbf{Extract 4} Shanghai is just the hotbed [-Attitude: Affect-security] of virus, bringing so much [Graduation: Force-quantification] trouble [-Attitude: Affect-happiness] to the whole [Graduation: Force-quantification] country. I'm not [Engagement: Heterogloss-dialogic contraction-disclaim] even able to take my wedding photos. I hate [-Attitude: Affect-happy] the virus carriers.

Extract 4 displayed the perceptions of people outside Shanghai towards the COVID outbreak. By comparing Shanghai to the hotbed of the virus, the poster expressed a negative attitude towards Shanghai, especially those who ran out of the city before the lockdown was enforced to avoid starvation and being quarantined \cite{zhang2023indoor}. The quantification of ``trouble'' and ``country'' highlighted the impact of COVID on society. The poster used ``hate'' to express their blame for the virus carriers. 

\textbf{Extract 5} Keep up scientific prevention and control! Adhere to the Dynamic Zero-COVID Policy! Everyone should do COVID testing! I believe [Engagement: Heterogloss-dialogic expansion-entertain] the epidemic in Shanghai can be controlled!

\textbf{Extract 6} The Chinese government faces more [Graduation: Force-intensifi-cation] challenges than Western governments to combat COVID. However [Engagement: Heterogloss-dialogic contraction-disclaim], Chinese people have the determination [+Attitude: judgment-social esteem-tenacity], confidence
[+Attitude: judgment-social esteem-capacity], and perseverance [+Attitude: judg-ment-social esteem-tenacity] to fight to the end! Believe in [+Attitude: Affect-security] the Party! Come on China!

Extracts 5 and 6 showed people's support of the COVID policy and the government. By using ``I believe'' in Extract 5, the poster presented their supportive stance on the government's policy. The adverb of degree ``more'' in Extract 6 represented the difficulty of containing the epidemic while the connective ``however'' invoked a contrary position \cite{Martin2005TheLO} that Chinese people were capable of combating COVID. The poster's belief and faith in the Chinese government were conveyed through the positive judgment of social esteem and positive affect. We noticed posters who strongly supported government policy were mostly non-Shanghai residents by checking their internet protocol (IP) locations \cite{zhang2022spatial}.

\textbf{Summary} During the Shanghai lockdown, there was a shortage of supplies. Complaints about the policy were commonly and strongly expressed in Shanghai. Some residents fled out of the city to avoid upcoming starvation and confinement, which was criticized by people in other cities for transmitting the virus. People outside Shanghai showed support for the COVID policy and the government, showing a contrasting difference compared with the attitude and sentiment within Shanghai. 

\subsubsection{Policy Termination}

The Chinese government loosened COVID restrictions in December 2022 and put an end to the three-year Dynamic Zero-COVID Policy. The lifting of COVID-19 restrictions led to a surge in infections and a shortage of medication in hospitals \cite{yan2023mobile}. Nevertheless, the public sentiment was positive.

\textbf{Extract 1} Just yesterday, someone in my family had a headache and went to the pharmacy to buy Ibuprofen. Unluckily [-Attitude: Affect-happy], he was told that it was out of stock.

\textbf{Extract 2} During the past three years of epidemic prevention and control, it was the medical workers who risked being infected to carry out COVID testing, quarantine, etc. After the restrictions were eased, it is also the medical workers who take all [Graduation: Force-quantification] responsibilities to handle the rising COVID cases. 

After the COVID restrictions were lifted, the problems of a surge of infections and drug shortage emerged. The lexical choice of ``unluckily'' in Extract 1 demonstrated the anxiety and unhappiness of the poster. The rapid spread of the virus posed pressure on the healthcare system and the word ``all'' in Extract 2 functioned as a quantifier for presenting the pressure of the medical works.

\textbf{Extract 3} In the past three years, we have embarked on the right [+Attitude: judgment-social sanction-propriety] path of coordinating epidemic prevention and control with economic and social development. Thanks to [+Attitude: Affect-satisfaction] the Party and the government for never giving up! Thanks to [+Attitude: Affect-satisfaction] the medical workers! Cheers to [+Attitude: Affect-happiness] Chinese people!

\textbf{Extract 4} Speaking of China's epidemic prevention and control measures, I think [Engagement: Heterogloss-dialogic expansion-entertain] it is world-leading [Graduation: Focus-sharpen] level [+Attitude: Appreciation-valuation]. Compared to other countries, 
the results of the epidemic prevention and control measures are quite [Graduation: Force-intensification] outstanding [+Attitude: Appreciation-reaction].

A positive attitude applauding the government, medical workers, and Chinese people was found in Extract 3. A positive judgment of the propriety (``right'') was attributed to the epidemic prevention and control measures in China. The poster used ``thanks'' three times to express their positive affectual reaction. In Extract 4, the prototypicality of China’s epidemic prevention and control measures was constructed as ``world-leading'' level, indicating the poster's positive appreciation of the policy. Furthermore, along with upscaled intensification (``quite''), positive evaluation of the results of the anti-epidemic measures  (``outstanding'') was situated within the global context by comparing with other countries.

\textbf{Extract 5} In the past, the control of the COVID-19 epidemic relied on restrictions. Although [Engagement: Heterogloss-dialogic contraction-disclaim] the economy suffered some losses, the results were extraordinary [+Attitude: Appreciation-reaction]. Now it depends on individual immunity and medications. I believe [Engagement: Heterogloss-dialogic expansion-entertain] our lives can be back to normal [+Attitude: Affect-happy] soon.

\textbf{Extract 6} Under this circumstance, many people feel more relaxed [+Attitude: Affect-security] than before. According to the latest anti-epidemic policies, COVID restrictions have been loosened. In this way, people's lives can be back to normal [+Attitude: Affect-happy].

Extracts 5 and 6 demonstrated public expectation and satisfaction for returning to a normal life via the employment of positive affect. The results of the three years of anti-epidemic measures were positively evaluated as ``extraordinary'' and the feelings toward the lift of the COVID restrictions were described as ``relaxed.''

\textbf{Summary} The lifting of COVID restrictions and the termination of the Dynamic Zero-COVID Policy led to a surge in COVID cases. Nevertheless, people expressed their satisfaction with the effect of three years of anti-pandemic measures and compared their success to other countries. The success was attributed to the government, medical workers, and Chinese people. Further, people expressed happiness for their lives returning to normal.
\section{Discussion}
By retrospectively analyzing public sentiments and attitudes toward the Dynamic Zero-COVID Policy in China as well as the epidemic prevention and control measures guided by the policy, we unveiled public perceptions of where and how the policy succeeded and failed. Below, we reflect on how public sentiments and attitudes toward the Dynamic Zero-COVID Policy changed over time, and implications for future epidemic prevention and control.

\subsection{Evolving Public Sentiments Toward the Dynamic Zero-COVID Policy}

By analyzing the sentiment of social media discussions concerning the Dynamic Zero-COVID Policy and the measures guided by the policy, we found a fluctuating trend in public sentiments. We selected key points, i.e., policy initialization and termination, sharp sentiment change, and lowest sentiment point, for a closer investigation through the lens of appraisal theory.  

At the start of the pandemic, little was known about its epidemiologic characteristics and transmission dynamics \cite{li2020early} and there was no vaccine to protect people from being infected. Thus, it was a serious challenge to both the government and the public. Previous studies showed that knowledge could affect both attitudes (e.g., perceived risk and efficacy belief) and practices (e.g., personal hygiene practices and social distancing) \cite{lee2021knowledge}, and the uncertainty about transmission paths of diseases may facilitate stigma against the disease \cite{stephen2023gender}. In the selected extracts, the common use of negative words reflected people's fear of the unknown virus. The frequent attitudinal attribution to experts and officials through the heteroglossic expression of quoted utterances demonstrated the public's dependence on and belief in the government and experts. When the real situation was in conflict with what the government and experts claimed, a strong sense of insecurity was displayed.

The public was generally supportive of the Dynamic Zero-COVID Policy in the early phase of the pandemic, comporting with existing research \cite{song2020public,xu2021public}. However, an attitudinal shift was noticed in March 2021 due to the continuous implementation of compulsory and strict epidemic prevention and control measures. A previous study found that people's attitudes were related to their perceptions of the epidemic situation and restrictions they were subject to \cite{sabat2020united}. As shown in the extracts, restrictions on mobility, e.g., the shutdown of public transportation, had a major impact on the everyday lives of people. The family members of the patients were discontent about the strict COVID testing requirements. Students in schools were dissatisfied with the restriction on their mobility and autonomy. The public's negative attitudes were reflected by the repeated application of negative judgment of the propriety of the policies. 

The sentiments reached the lowest point during the heightened restriction during the Shanghai lockdown \cite{zhang2023expressed}. People felt stressed out and frustrated owing to the lack of food, income, medicine, and other necessities \cite{zhai2023lockdown}. As presented in the extracts, adjectives expressing negative attitudes were constantly used along with upscaled intensification, reflecting Shanghai residents’ criticism of the policy. Blames were placed on the people who left Shanghai and brought the virus to other provinces. An examination of the IP locations of the posts disclosed that non-Shanghai residents were generally more positive than Shanghai residents by using the positive judgment of social esteem, which is consistent with the prior literature \cite{zhang2023expressed}. 

After the Dynamic Zero-COVID Policy was lifted in December 2022, people expressed their relief and expectation for a normal life despite concerns about the rise of COVID cases. Most posts were positive appreciation and judgment of the government's COVID policy and measures, which coincides with a past investigation revealing that Chinese people showed more trust in the government, more positive attitudes towards prevention measures, and more compliance with COVID-19 prevention guidelines compared to people in the US \cite{wang2022need}.

\subsection{Implications for Future Pandemics}
In recent years, the whole world has witnessed many public health emergencies that greatly influence human society, including SARS in 2003 \cite{huang2004sars}, avian influenza in 2004 \cite{chen2009h5n1}, H1N1 influenza in 2009 \cite{koh2010wen}, H7N9 influenza in 2013 \cite{kageyama2013genetic}, West Africa Ebola virus in 2014 \cite{gomes2014assessing}, and COVID-19 in 2020 \cite{ciotti2020covid}. An effective response to and governance of public opinions during epidemic situations are important in social emergency management \cite{zhao2022modeling}. Proper emergency orientation is not only conducive to the incident's resolution but also to the construction of a harmonious society \cite{lian2022public}. Here, we propose implications for more effective epidemic prevention drawing on the lessons learned during COVID in China.

\subsubsection{Transparent Information Sharing Between Governments and The Public}
As observed in the extracts, the Chinese government was hesitant to share the information with the public and ignored the significance of transparent information disclosure at some point \cite{lian2022public}. The public was in panic after they found that the real situation conflicted with the statements of the government. During the pandemic, people are in urgent need of obtaining the latest and objective information from government agencies \cite{lian2022public}. 
Policymakers and governments are expected to adopt effective strategies to provide accurate information to secure a sufficient level of trust and confidence from the public \cite{sabat2020united}.

\subsubsection{Public Sentiment and Attitude: Two Intertwining Signals}
Our analysis revealed a correlation between public sentiment at the macro level and people's attitudes toward the Dynamic Zero-COVID Policy at the micro level. For example, during the Shanghai lockdown, public sentiment reached its lowest point. A closer observation of social media discussions revealed people's lived experience of anxiety, panic, and frustration of the lockdown as well as COVID policy and measures in general. It is important to combine sentiment analysis and fine-grained attitude/perception analysis to understand the needs of the public and inform corresponding policy \cite{whitelaw2005using}.

\subsubsection{Social Media Censorship}

We noticed severe censorship of Weibo posts \cite{bamman2012censorship} regarding COVID and the related policies. Only 18 posts were found on December 31st, 2019 using COVID-related keywords, which was far from the reality. In the process of data collection, we found that many original posts that had been forwarded/retweeted were deleted. Prior research also indicated that a broad range of COVID-19-related content was censored on Chinese social media at the early stage of the epidemic \cite{ruan2020censored}.
The censorship and surveillance of public discourse on epidemics and crises are not helpful for crisis resolution due to their potential to decrease political trust and facilitate the future accessibility of uncensored information regarding the government \cite{chang2022covid}.

\subsection{Limitations and Future Work}
There are several limitations of our study. First, the sample size of our collected Weibo posts is relatively small (n=2,358). Thus, we do not claim the generalizability of the public sentiments and attitudes toward the Dynamic Zero-COVID Policy to the whole Chinese population. Rather, we aimed to provide a retrospective snapshot of how public sentiments and perceptions evolved during the implementation of the policy, and present people's lived experiences during COVID. Second, Weibo posts are susceptible to censorship, further degrading the representativeness of the data. Future work could leverage a large-scale survey study to understand the censored emotions towards the policy.
\section{Conclusion}
Through a combination of large-scale sentiment analysis and in-depth discourse analysis, we unveil the evolving trends of public sentiments and perceptions of the Dynamic Zero-COVID Policy in China. In the initial stages of the epidemic, the public’s emotions were overall pessimistic due to the widespread panic stemming from the uncertainty of the virus. After three years of implementing anti-epidemic measures, the Dynamic Zero-COVID Policy has garnered positive evaluations from the public, who expressed their gratitude towards the government, the medical workers, and the Chinese people. The public sentiment underwent a shift in March 2021 as a result of the prolonged restrictions imposed on various aspects of people’s lives and people's dissatisfaction with certain unreasonable measures. The public sentiment reached its lowest point during the lockdown of Shanghai in 2022 when frustration and panic swept the city. Based on the findings, we propose implications for future epidemics, such as transparent information sharing by governments and free speech on social media. 

\bibliographystyle{splncs04}
\bibliography{covid}

\end{document}